\documentclass[aps,superscriptaddress]{revtex4}
\usepackage{amsmath,amssymb}
\usepackage{graphics,graphicx}
\usepackage{dcolumn,bm}

\newcommand{\sinp}{\affiliation{Centre for Applied
Mathematics and Computational Science
and Theoretical Condensed Matter Physics Division,\\
Saha Institute of Nuclear Physics, 1/AF Bidhannagar, Kolkata 700 064, India.}}
\newcommand{\isik}{\affiliation{Economic Research Unit,
Indian Statistical Institute, Kolkata Centre,
203 Barrackpore Trunk Road, Kolkata 700108, India.}}
\newcommand{\ictp}{\affiliation{Condensed Matter and Statistical
Physics Section,\\
The Abdus Salam International Centre for Theoretical Physics,
Strada Costiera 11, Trieste I-34014, Italy.}}

\begin{document}
\title{The Kolkata Paise Restaurant Problem and Resource Utilization}
\author{Anindya-Sundar Chakrabarti}
\email[Email: ]{x.econo@gmail.com}
\isik
\author{Bikas K. Chakrabarti}
\email[Email: ]{bikask.chakrabarti@saha.ac.in}
\sinp \isik
\author{Arnab Chatterjee}
\email[Email: ]{achatter@ictp.it}
\ictp
\author{Manipushpak Mitra}
\thanks{Corresponding author}
\email[Email: ]{mmitra@isical.ac.in}
\isik


\begin{abstract}
We study the dynamics of the ``Kolkata Paise Restaurant problem''. 
The problem is the following: In each period, $N$ agents have to 
choose between $N$ restaurants. Agents have a common ranking of the 
restaurants. Restaurants can only serve one customer. When more than 
one customer arrives at the same restaurant, one customer is chosen 
at random and is served; the others do not get the service. We first introduce the one-shot versions of the Kolkata Paise Restaurant problem which we call one-shot KPR games. We then study the dynamics of the Kolkata Paise Restaurant problem (which is a repeated game version of any given one shot KPR game) for large $N$. For statistical analysis, we explore the long time steady state behavior. In many such models with myopic agents we get under-utilization of resources, that is, we get a lower aggregate payoff compared to the social optimum. We study a number of myopic strategies, focusing on the average occupation fraction of restaurants.  
\end{abstract}
\maketitle 

\section{Introduction}
\noindent Following the earlier version of the Kolkata Restaurant Problem \cite{krp-orig} as a generalization of the El Farol Bar Problem \cite{arthur1994} and the consequent Minority Game Problem \cite{challet2005}, we introduce a very simple version of the game: the Kolkata Paise Restaurant (KPR) problem, which is a repeated game with no mutual interaction. The set $N$ of agents (restaurants) can be macroscopically large and agents have a common preference ranking across the $N$ restaurants. We first introduce one-shot KPR games. Any given one-shot KPR game is repeated each period in the KPR problem. For any one-shot KPR game the set of pure strategy Nash equilibria coincides with the set of socially (Pareto) efficient outcomes. By socially efficient outcomes we mean outcomes where each agent goes to a separate restaurant and each restaurant gets exactly one agent to serve. If agents are rational and $N$ is small, one can sustain any socially efficient outcome in equilibrium of the repeated game version of the KPR problem (see \cite{kandori}). However, we assume that $N$ is large and hence an agent cannot rely on other agents actions and therefore what matters to any agent is the past collective configuration of actions and the resulting average utilization of restaurants. Assuming very `simple' stochastic strategies based on history and past individual experiences and assuming no mutual interaction, we observe that the KPR problems are `intrinsically dynamic' in the sense that, the system in general never converges to any of the states and most of the time, the system passes through socially inefficient states (for a special example of converging dynamics, see Sec.~\ref{os}). 
Our main objective is to analyze the statistics of the fraction of restaurants utilized per day (utilization fraction $f$) associated with various stochastic strategies. In particular, we investigate some of the optimally utilized states of the system, the average occupation (utilization) fraction of the restaurants on any day and their fluctuations.

\section{The problem}
\noindent
Let there be $N$ restaurants in Kolkata and assume that each can accommodate a fixed number of customers, normalized to unity. Let there be exactly $N$ customers or agents in the city. Though each of the restaurants costs the same, we assume that there is a common preference ranking of these restaurants to the agents. If, in a day,  there is more than one agent (prospective customer) arriving in any particular restaurant, one of them is chosen randomly and the rest do not get the service. {\it In Kolkata, there happened to exist  very cheap and fixed rate ``Paise Restaurants" (also called ``Paise Hotels''), popular among the daily laborers in the city. During lunch hours, the laborers used to walk down (to save the transport costs) to any of these restaurants and would miss the lunch if they arrived at a restaurant where their number is more than the capacity of the restaurant for such cheap lunch. Walking down to the next restaurant would mean failing to report back to the job in time! Paise means the smallest Indian coin and there were indeed some well known rankings of these restaurants as some of them would offer more tastier items compared to the others}.

Herd behavior and herd externality (see \cite{orlean,banerjee}) cannot explain the omnipresence of social inefficiency for the KPR problem since we assume that there is no ambiguity among agents over the ranking of the restaurants. The KPR problem is similar to the Minority game since it punishes herd behavior and promotes diversity (see \cite{challet2005}). The KPR problem is different from the Minority game in the following ways: (a) while the minority game is a simultaneous move {\it two choice} problem, the KPR problem is a simultaneous move {\it many choice} problem and (b) for the KPR problem there is a common preference ranking of these restaurants and this sort of ranking is absent for the minority game.

We assume that while deciding on the choice of the restaurant, there is no mutual interaction among the agents, and that they  decide simultaneously on the basis of own past ``experiences'' and complete history. There are  $N^N$ possible outcomes of which most are socially inefficient in the sense that there exists at least one agent not getting lunch and equivalently, there is at least one restaurant without a customer. There are exactly $N!$  (that is, a fraction exp$[-N]$ as $N \to \infty$) outcomes that correspond to socially efficient utilization of the restaurants where each agent ends up in a different restaurant. Given that all agents pay the same price (though rankings are different), in general a socially efficient outcome is not individually optimal in this repeated set up. Therefore, even when there is a realization of the efficient outcome in one period, it gets destabilized in the next period and the whole system breaks down to any of the infinitely many ($\exp[N]$; $N\to \infty$) socially inefficient states. Hence there can be no fixed point result for the problem and the game continues dynamically forever. The system moves to an inefficient state characterized by overcrowding at higher ranked restaurants (absence of any ``absorbing'' state). It is precisely this generic deviation from the efficient outcome that motivated us to investigate some of the optimally utilized states of the system and the average occupation (utilization) fraction of the restaurants on any day. Therefore, our attempt is to analyze different versions and modifications of the KPR problem using various deterministic and stochastic strategies.
  
\section{One-shot KPR game}\label{os}
\noindent
How does one capture the features of the Kolkata Paise Restaurant problem of {\it any given day}? That is, the situation where on any given day
\begin{enumerate}
\item[(a)] each agent goes for lunch to any one of the $N$ restaurants,
\item[(b)] depending on the number of agents in each restaurant only one agent is selected randomly and allowed to have lunch, and 
\item[(c)] the agent derives a utility from having lunch and this utility level depends on the rank of the restaurant and if the agent does not get lunch then his utility is zero on that particular day.
\end{enumerate}

One can use the tools available in the game theory literature to model  the Kolkata Paise Restaurant problem {\it of any given day}  as a one-shot game. We now introduce a general one-shot {\it common preference  restaurant game} and identify the conditions under which it represents  the {\it one-shot KPR game}. Let $\mathcal{N}=\{1,\ldots,N\}$ be the set of agents $(N<\infty)$ and let the vector $u=(u_1,u_2,\ldots,u_N)\in \Re^N$ represent the utility (in terms of money) associated with each restaurant which is common to all agents. Assume without loss of generality that $0<u_N\leq \ldots \leq u_{2}\leq u_1$. These three things (that is, $\mathcal{N}$, $u=(u_1,\ldots,u_N)$ and the ranking $0<u_N\leq \ldots \leq u_2\leq u_1$) are used to define the one-shot common preference restaurant game. Formally, $G(u)=(\mathcal{N},S,\Pi)$ represent a one-shot common preference restaurant game where $\mathcal{N}=\{1,\ldots,N\}$ is the set of agents (players), $S=\{1,\ldots,N\}$ is the (common) strategy space of all agents where a typical strategy $s_i=k$ denotes the strategy that the $i$-th agent's strategy is to go to the $k$-th restaurant. In other words, each agent on any given day selects any one of the $N$ restaurants numbered $1$ to $N$ and if any agent $i$ decides to go to restaurant $k\in \{1,\ldots,N\}$ then we represent it as $s_i=k$. We denote by $\Pi=(\Pi_1,\ldots,\Pi_N)$ the expected payoff vector. In particular, $\Pi(s)=(\Pi_1(s),\ldots,\Pi_N(s))$ is the expected payoff vector associated with any strategy combination $s=(s_1,\ldots,s_N)\in S^N$ where player $i$'s payoff is $\Pi_i(s)=\frac{u_{s_i}}{n_i(s)}$ and $n_i(s)=1+|\{j\in \mathcal{N}\setminus \{i\}\mid s_i=s_j\}|$.  Therefore, the strategy combination $s=(s_1,\ldots,s_N)$ represents the selection of restaurants made by the agents and the expected payoff that an agent $i\in \mathcal{N}$ gets depends not only on which restaurant he has selected but also on how many other agents have selected the same restaurant. The number of agents selecting the same restaurant as that of agent $i$ under the strategy combination is given by $n_i(s)$. Since we have assumed that the restaurant selects one agent randomly from the set of $n_i(s)$ agents it follows that the expected payoff of agent $i$ given the strategy combination $s=(s_1,\ldots,s_N)$ is $\Pi_i(s)=\frac{u_{s_i}}{n_i(s)}$.  For each selection of a vector $u=(u_1,u_2,\ldots,u_N)$ with the property that $0<u_N\leq \ldots \leq u_2\leq u_1$ we associate a common preference restaurant game $G(u)=(\mathcal{N},S,\Pi)$. We now introduce added restrictions on the common preference restaurant games to represent the one-shot KPR games. 

The first feature of the KPR problem is that agents have a common 
preference ranking over the restaurants. If all restaurants are 
identical to the agents (that is, if $u_1=\ldots,u_N$) then the problem is trivial. Hence to introduce the problem in a non-trivial way it is reasonable to assume that for the one-shot KPR game $u_1\not =u_N$. The next feature of the KPR problem is that agents prefer getting lunch to not getting lunch. To capture this feature we assume that the payoff specification of a one-shot KPR game must be such that $u_{k'}>\frac{u_k}{2}$ for all $k,k'\in \{1,\ldots,N\}$ and given $0<u_N\leq \ldots \leq u_2\leq u_1$ and $u_1\not =u_N$ this is equivalent to assuming that $u_N > \frac{u_1}{2}$. Intuitively, if $u_N>\frac{u_1}{2}$ then even the person who was at the lowest ranked restaurant, would not be willing to alter his strategy. Therefore, {\it if a one-shot common preference restaurant game $G(u)=(\mathcal{N},S,\Pi)$ is such that $0<u_N\leq \ldots \leq u_2\leq u_1$, $u_1\not =u_N$ and $u_N > \frac{u_1}{2}$ then it is a one-shot KPR game}. 

We say that a strategy combination $s=(s_1,\ldots,s_N)$ leads to socially efficient (or Pareto efficient) outcome if and only if each agent goes to different restaurants that is $s_i\not =s_j$ 
for all $i,j\in \mathcal{N}$, $i\not =j$. A strategy combination 
$s=(s_1,\ldots,s_N)$ is a pure strategy Nash equilibrium if no agent has an incentive to deviate from his existing strategy $s_i$ given the strategy of the other players.  For the above mentioned version of the one-shot KPR game {\it the set of pure strategy Nash equilibria coincides with the set of socially efficient allocations}. This is formally established in the appendix. 

In this section we have introduced the one-shot common preference restaurant games and identified conditions under which these games also represent the one-shot KPR games. The key features that separate out a one-shot KPR game from other non-KPR one-shot common preference restaurant games are non-triviality ($u_1\not=u_N$) and the restriction $u_N > \frac{u_1}{2}$ that ensures that it is always in the interest of any agent to choose an unoccupied restaurant.

\section{Simple stochastic strategies and utilization statistics}\label{sec:stochgm}
\noindent
It is well known from the repeated game theory literature that if {\it agents can coordinate their actions} then any pure strategy Nash equilibrium outcome of the one-shot KPR game (which is also a socially efficient outcome) can be sustained as an equilibrium outcome of KPR problem (the repeated one-shot KPR game) with appropriate punishment schemes (see \cite{kandori}). However, sustaining  such equilibrium in repeated play of the one-shot KPR game would require that {\it set of agents is small so that they can coordinate their actions}. Since expecting such coordinated actions from the agents is unrealistic for the KPR problem (where $N$ is large) we assume that with repetition of a one-shot KPR game agents work out their strategies without coordination.

Following Brian Arthur's tradition we assume that agents reason inductively (see \cite{arthur1994}). By inductive reasoning we mean that agents learn from past experiences, discard their old strategy for something new if it performs badly and continues with the old strategy if it performs well. However, our objective in this paper is not to find out strategies that survive this process of inductive reasoning but to analyze the statistics of the fraction of restaurants utilized per day associated with each such strategy. We assume that all agents take up some (common) stochastic strategy independent of each other. By stochastic strategies we mean that agents choose their strategies in a probabilistic way. At any step this choice of probabilities can depend on the outcome of the previous period.  In each case we consider the limit $N \to \infty$ and analyze the statistics of the fraction $f$ of the restaurants utilized per period. The specific one-shot KPR game $\bar{G}(u)$ that we use is the one where $0<u_N<\ldots<u_1$ so that agents have a strict ranking and $\frac{u_1}{2}<u_{N}$. 

\subsection{No learning (NL) strategy} {\it On each day $t$ an agent selects any one restaurant with equal probability. There is no memory or learning and each day the same process is repeated.}

\begin{figure}
\includegraphics[width=12.0cm]{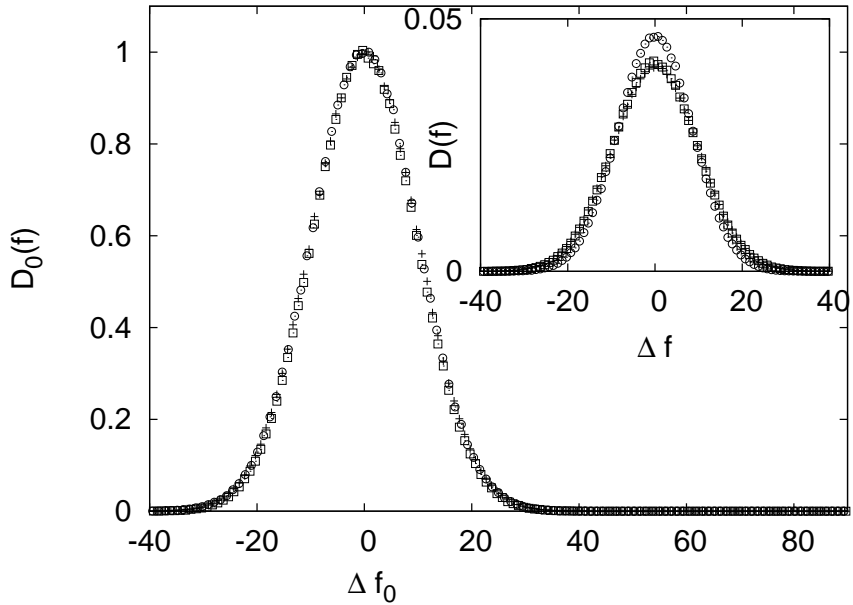}
\caption{\label{fig:1}
Numerical simulation results for the normalised 
(by the respective half widths)
distribution $D_0(f)$ of the
fraction $f$ of people getting lunch on any day (or fraction
of restaurants occupied) against $\Delta f \equiv f-\bar{f}$
 where $\bar{f}$ denote the
average occupancies in each cases.
The inset shows the bare distributions $D(f)$ against $\Delta f$.
All the numerical simulations have been done
for $N=1000$ agents for $10^6$ time steps.
$\bar{f} \simeq 0.63$ in case NL,
$\bar{f} \simeq 0.43$ in case LL(1) and
$\bar{f} \simeq 0.77$ in case OPR, as estimated
analytically (Eqns.~(\ref{eq:Pm0}), (\ref{eq:fbart}) and (\ref{eq:ft+1_2})).
The data points are ($+$),  ($\Box$) and ($\bigcirc$) for cases NL, LL(1) 
and OPR respectively.
}
\end{figure}

If any restaurant is chosen by $m \hspace{.01in}(>1)$ agents on a particular day $t$, the restaurant chooses one of them randomly, and the rest $m-1$ agents go without any lunch that day. Let there be $\lambda N$ agents ($\lambda=1$ in KPR problem) choosing randomly from the set of $N$ restaurants. Then the probability $p$ of choosing any restaurant is $1/N$. The probability $P(m)$ that any restaurant is chosen simultaneously by $m$ agents is therefore given by Poisson distribution
\begin{eqnarray}
\label{eq:poisson_mm}
P(m) &=& \left( \begin{array}{c} \lambda N \\ m \end{array}\right)
p^m (1-p)^{\lambda N -m}; \ \ p=\frac{1}{N}  \nonumber \\
&=& \frac{\lambda^m}{m!} \exp({-\lambda}) \ \ {\rm as} \ \ N \to \infty
\end{eqnarray}
In other words, the fraction of restaurants not chosen any day is
given by $P(m=0) = \exp(-\lambda)$, giving the average fraction of
restaurants occupied
\begin{equation}
\label{eq:Pm0}
\bar{f}= 1- \exp(-\lambda) \simeq 0.63 \ {\rm for} \ \lambda=1,
\end{equation}
in the KPR problem. The distribution $D(f)$ of the fraction utilised any day will be a
Gaussian around the average given above. See the simulation results in Figure ~\ref{fig:1} for $N=1000$.

\subsection{Limited learning (LL) strategy} 
We consider two types of learning strategies and show that in both the 
cases the utilization fraction $\bar{f}$ goes down compared to the no learning case. 
\begin{enumerate}
\item[{\bf LL(1)}] {\it If an agent gets lunch on day $t$ then the agent goes to the best restaurant on day $t+1$. Moreover, if on some day $t$ the agent goes to best restaurant and fails to get lunch then, on day $t+1$, the agent selects a restaurant from the remaining $N-1$ restaurants with equal probability.} 
\end{enumerate}
When $f_t N$ number of agents get their lunch on any day $t$ and decide to get their next day ($(t+1)^{\rm th}$ day) lunch in the best ranked restaurant, then only one of them will get into the highest ranked restaurant and others will not get their lunch. The rest $N - f_t N$ agents will try independently for the remaining $N-1$ restaurants, following the same strategy as in the no learning case. Then the recursion relation will simply be 
(following Equation ~(\ref{eq:Pm0})):
\begin{equation}
\label{eq:fbart}
f_{t+1} = 1-\exp(-\lambda_t); \ \lambda_t=1-f_t
\end{equation}
giving the fixed point fraction $\bar{f} \simeq 0.43$.
This again compares well with the numerical simulation results (for $N=1000$) and shows a Gaussian distribution $D(f)$ around the above value of $\bar{f}$ as in Figure ~\ref{fig:1}.

\begin{enumerate}
\item[{\bf LL(2)}] {\it If an agent reaches the $k$-th ranked restaurant on 
day $t$ then the agent's strategy is the following: 
(a) if the agent gets lunch on day $t$ then, on day $t+1$, 
the agent attempts to select a restaurant from the rank set 
$\{k-1,\ldots,1\}$ with equal probability and (b) if the agent does not get 
lunch on day $t$ then, on day $t+1$, the agent attempts to select a 
restaurant from the set $\{k,\ldots,N\}$ with equal probability.}
\end{enumerate}

For this strategy we get the fixed point fraction $\bar{f}\simeq 0.5$ and it shows a Gaussian distribution around $\bar{f}$. Hence in this case also there is no improvement compared to the NL case. If we can somehow introduce repetition  in the strategy of the agent then we get an improvement in the utilization fraction.
In fact, since the reverse choice problem, where for (a) the choice
would be $\{k,\ldots,N\}$ and for (b) would be $\{k-1,\ldots,1\}$,
is identical, we get $\bar{f}=1-\bar{f}$, giving $\bar{f}=0.5$.

\subsection{One period repetition (OPR) strategy} 
\label{ss:opr}
{\it If an agent gets lunch on day $t$ from restaurant $k$ then the agent goes back to the same restaurant on day $(t+1)$ and competes for the best restaurant (restaurant $1$) on day $(t+2)$. If the agent fails to get lunch on day $t$ then the agent tries for a restaurant on day $(t+1)$ which was vacant on day $t$, using the same stochastic strategy as in the NL case.}

The fraction $f_t$ on any day $t$ of the people occupying any restaurant consists of two parts: fraction $x_{t-1}$
of the people already continuing in their randomly selected restaurant
the earlier day and the fraction $x_t$ of people who have chosen today.
As such, using Equation ~(\ref{eq:Pm0}),
\begin{equation}
\label{eq:ft_2}
f_t = x_{t-1} + (1-x_{t-1})[1-\exp(-1)].
\end{equation}
Since the fraction $x_t$ chosen today is given by the fraction $\bar{f}$
in Equation ~(\ref{eq:Pm0}) where $N(1-x_{t-1})$ left out agents are choosing
randomly out of $N(1-x_{t-1})$ vacant restaurants.
The next day, the fraction $f_{t+1}$ will therefore be given by
\begin{equation}
\label{eq:ft+1}
f_{t+1}=(1-x_t)(1-\exp(-1))+[1-(1-x_t)(1-\exp(-1))][1-\exp(-1)].
\end{equation}
If we assume that $f_{t+1}=\bar{f}=f_t$ and $x_t = x$ at the fixed point, then 
equating (\ref{eq:ft_2}) and (\ref{eq:ft+1})
\begin{equation}
\label{eq:ft+1_2}
x = [1-(1-x)(1-\exp(-1))](1-\exp(-1)), 
\end{equation}
or $x \simeq 0.38$, giving $\bar{f} \simeq 0.77$,
as can be seen in the numerical simulation results in Figure ~\ref{fig:1}. The fluctuations in the occupation density or utilization fraction $D(f)$ is again Gaussian around the most probable occupation fraction derived here. Therefore, with OPR strategy we have an increase in $\bar{f}$ compared to the NL strategy suggesting that if the strategies are such that an agent, after getting lunch on a particular day, decides to go to the same restaurant next day we have an improvement in utilization fraction.

\noindent \textbf{Remark} 
The OPR strategy is more consistent when the one-shot common preference restaurant game $G(u)$ is such that $u=(u_1=\ldots =u_N)$. It may be noted that if all the restaurants are identical then the dynamics always converge to an absorbing state characterized by social efficiency (with Pavlov's {\bf win-stay, lose-shift (WS-LS)} strategy (see \cite{Nowak})). The reason being the following: even if everybody starts by random choices among all the $N$ restaurants and, initially (say, on day $t$), several restaurants get more than one prospective customer and choose one of them, on day $t+1$, the agents not getting lunch on day $t$ will go to a restaurant that was unoccupied on day $t$ and the agents getting lunch on day $t$ will continue to go to the same restaurant. Soon (before, $N+1$ days), each agent  finds one restaurant to get the lunch and the dynamics effectively stops there.\label{trivial}

In the next case we think of a strategy where the queue length in a restaurant on day $t$ either acts as an inducing device or acts as a rationing device for the strategy to be adopted by an agent on day $t+1$.  

\begin{figure}
\includegraphics[width=12.0cm]{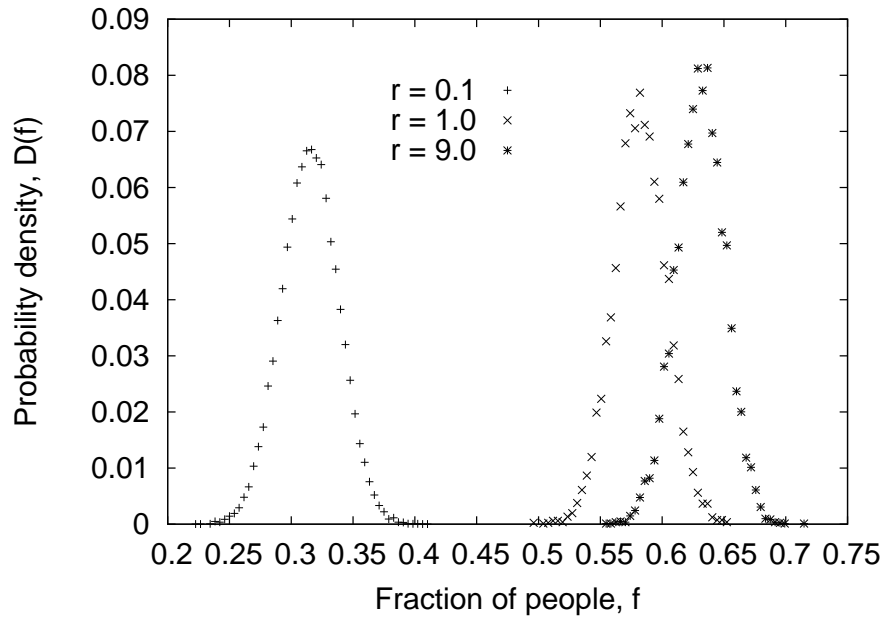}
\caption{\label{fig:2}
Numerical simulation results for the distribution $D(f)$ of the
fraction $f$ of people getting lunch any day (or fraction
of restaurants occupied on any day) against $f$, for
$r=0.1, 1.0$ and $9.0$ for case FC.
All the numerical simulations have been done
for $N=512$ agents for $10^6$ time steps.
}
\end{figure}

\subsection{Follow the crowd (FC) strategy} \label{ss:fc}
\noindent
{\it The probability of selecting a restaurant $k$ on day $t+1$ depends on the queue length of agents in restaurant $k$ on day $t$.} 

We assume that the probability $p_k(t+1)$ of choosing $k^{\rm th}$ restaurant on $(t+1)^{\rm th}$ day is given by 
\begin{equation}
p_k(t+1) = \frac{ q_k(t) + r}{N (1 + r)}
\end{equation}
where $q_k(t)$ is the queue length (or equivalently, the number of people who 
arrived at the $k^{\rm th}$ restaurant on day $t$). $r$ is a parameter that interpolates between the random ($r \to \infty$, independent of history) and condensation ($r \to 0$, strongly dependent only on history) limits. The representative histograms are shown in the Figure 2, where the most probable fraction $\bar{f}$ of occupied restaurants  converge to a value 0.63 (as obtained earlier)
as $r \to \infty$. It is also seen that the probability of a queue length
$q$ of any restaurant goes as $q^{-1}$ for $r \to 0$ (exponentially decaying
otherwise). This implies that while the utilization fraction has a Gaussian distribution (with $\bar{f}=0.63$) the queue length distribution on any arbitrary day $t$ has a power law distribution if $r \rightarrow 0$ and it is exponentially decaying otherwise.

If the agents follow {\bf avoid the crowd (AC)} strategy  that is if $p_k(t) \to 1 - p_k(t)$ (people choosing to avoid the last day's crowded restaurants), the most probable restaurant utilization fraction $\bar{f} \simeq 0.63 $ is independent of the value of $r$ $(> 0)$. Here, no power law for the queue length was observed for any value of $r$. The observations of this subsection are similar to the Hospital waiting list size distribution (see \cite{frekleton2001}, \cite{smethurst2001}) though the Hospital problem is significantly different from the KPR problem (for example, the  absence of possibilities of iterative learning and the presence of heterogeneity of choices in the Hospital waiting list problem).
 
\section{Limited queue length and modified KPR problem} \label{sec:lql}
\noindent
Till now we had assumed that on any day $t$ one can have potentially all the $N$ agents turning up in one particular restaurant. If the arrival of the agents were sequential then all agents turning up in one restaurant tantamount to saying that queues of all sizes are allowed to form in any restaurant and the restaurant then picks up any one agent to serve. All other agents have to go back without having lunch on day $t$. Here we are assuming that formation of queue of all sizes is allowed. What if all restaurants have limited capacity to accommodate agents in a queue and that an agent gets the  option to move to another restaurant (in the same day) if the agent arrives at a restaurant that has reached its queue limit? Before concluding we consider this particular modification of the KPR problem since it has important implications about the robustness of our occupancy statistics $\bar{f}$.  

\noindent
\textit{A modified KPR problem:} 
On each day the restaurants are filled up sequentially. We also assume that {\it each restaurant allows for a queue length $q^*$, so that the $(q^*+1)^{\rm th}$ person arriving at the $k^{\rm th}$ restaurant has to search for another restaurant with a queue length less than $q^*$}. When all the customers have arrived, one of the $q_i\leq  q^*$ customers gets lunch at the $k^{\rm th}$ restaurant, while the $q_i-1$ number of prospective customers do not get any lunch that day. In terms of the dynamics, the most probable restaurant utilization fraction $\bar{f} = 1$ occurs for $q^*=1$. We find that for $q^* = 2$, the fraction $\bar{f}$ {\it drastically falls} to $0.68$. Also for $q^*\rightarrow N$, $\bar{f}=0.63$ as discussed and derived earlier. Figure 3 shows the result for different values of $q^*$. Note that $\bar{f}$ approaches the value for the random occupancy case as $q^*$ becomes large. Thus, except for the special case where $q^*=1$, for all other cases we get conclusions that are closer to the KPR problem derived in the previous section. 

\begin{figure}
\includegraphics[width=12.0cm]{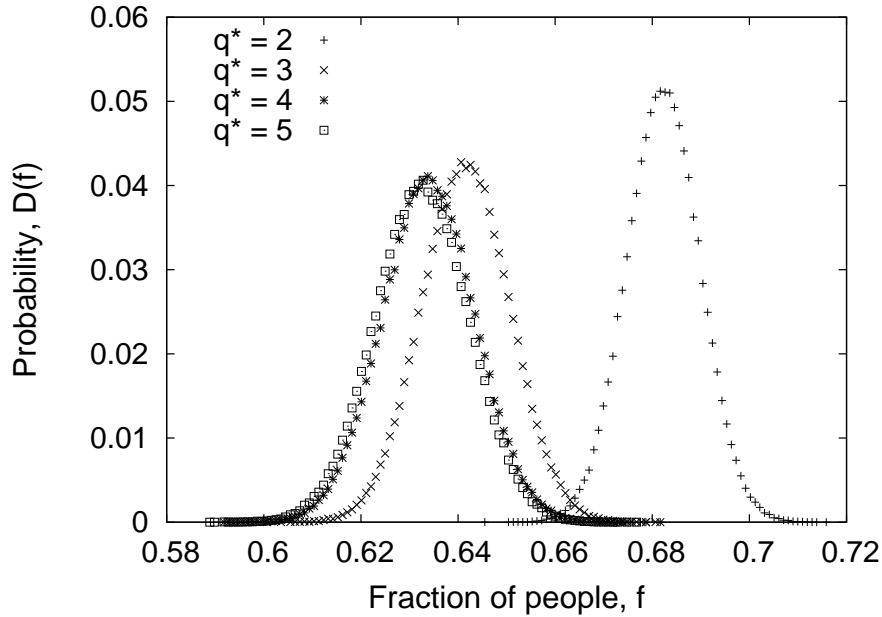}
\caption{\label{fig:th}
Numerical simulation results for the distribution $D(f)$ of the
fraction $f$ of people getting lunch any day (or fraction
of restaurants occupied on any day) against $f$, for
$q^*= 2, 3, 4, 5$  for the modified KPR problem.
All the numerical simulations have been done
for $N=1024$ agents for $10^6$ time steps.
}
\end{figure}

\section{Discussion and Conclusion}
\noindent
For the general KPR problem we have a perpetual dynamics characterized by large fluctuations in the measure of daily misuse having a Gaussian distribution of utilization fraction $f$ around the average $\bar{f}$ where $\bar{f}$ differs for different strategies. We have argued this point by showing that the average utilization fraction $\bar{f}$ 
with no learning and other simple stochastic strategies could be analytically calculated and the distributions are seen to be Gaussian. For the ``follow/avoid the crowd'' strategies we observe the statistics of the number of people {\it going without lunch} to have power law fluctuations (see also \cite{arthur1994}, 
\cite{challet2005}, \cite{frekleton2001}, \cite{smethurst2001} for similar behavior).

Our main findings in terms of utilization fraction comparisons across different strategies are: 
\begin{enumerate}
\item[(a)] If we have a one-shot KPR game then any pure strategy Nash equilibrium leads to a socially efficient outcome, as shown in Sec.~\ref{os}, and hence we have full capacity utilization. However, if we have a KPR problem (which is a repeated game where any given one-shot KPR game is repeated) and agents cannot coordinate their action then we do not have full capacity utilization for many (common) stochastic strategies. In such a situation utilization fraction under different stochastic strategies is of prime importance. 
\item[(b)] The utilization fraction is higher on an average with no learning strategy in comparison to some limited learning strategies 
(Sec.~\ref{sec:stochgm}). 
\item[(c)] The statistics for utilization fraction can improve with a strategy where each agent, after getting lunch from a restaurant on day $t$, goes back to the same restaurant on day $(t+1)$ (ignoring the ranking of the restaurants) (Sec.~\ref{ss:opr}).
\item[(d)] Following (avoiding) the crowd strategy cannot improve on the statistics of utilization fraction in comparison to the no learning case (Sec.~\ref{ss:fc}). 
\item[(e)] Our findings about the statistics of utilization fraction are robust since we have shown that by adding different queue limits to the KRP problem the conclusions are similar to our findings except for the very special case where the queue limit is unity (Sec.~\ref{sec:lql}).  
\end{enumerate} 

\vspace{.2in}

\noindent {\bf Acknowledgment:} We are grateful to Satya Ranjan Chakravarty, Pradip Maiti, Pradeep Kumar Mohanty, Abhirup Sarkar and Arunava Sen for some useful discussions and suggestions. AC thanks CAMCS, Saha Institute of Nuclear Physics for supporting
a visit from ICTP, during which the work started.
AC was supported by the ComplexMarkets E.U. STREP project  
516446 under FP6-2003-NEST-PATH-1.

\section{Appendix}
\noindent
A strategy combination $s\in S^N$ for the one-shot common preference restaurant game $G(u)=(\mathcal{N},S,\Pi)$ leads to {\it socially (Pareto) efficient} outcome if and only if $n_i(s)=1$ for all $i\in N$. In other words, given any one-shot common preference restaurant game $G(u)$ a strategy combination leads to a socially efficient outcome if and only if each agent goes to a different restaurant so that the sum of utilities of the agents is $\sum_{k=1}^Nu_k$. Let $s_{-i}$ denote the strategy combination of all but player $i$. A strategy combination $s^*\in S^N$ is a {\it pure strategy Nash equilibrium} of the one-shot common preference restaurant game $G(u)=(\mathcal{N},S,\Pi)$ if for all $i\in \mathcal{N}$, $\Pi_i(s^*)\geq \Pi_i(s_i,s^*_{-i})$ for all $s_i\in S$. For a typical one-shot common preference restaurant game $G(u)=(\mathcal{N},S,\Pi)$, let $SE(G(u))=\{s^*\in S^{\mathcal{N}}\mid n_i(s^*)=1\hspace{.025in}\forall \hspace{.025in} i\in \mathcal{N}\}$ denote the set of all strategy combinations that lead to socially efficient allocations and let $PN(G(u))=\{s^*\in S^{\mathcal{N}}\mid \hspace{.025in} \forall \hspace{.025in} i\in \mathcal{N}, \Pi_i(s^*)\geq \Pi_i(s_i,s^*_{-i}) \hspace{.025in} \forall \hspace{.025in} s_i\in S\}$ denote the set of all pure strategy Nash equilibria. Let $\bar{G}(u)$ represent any one-shot KPR game we introduced in Sec.~\ref{os}, that is any one-shot common preference restaurant game with the property that $0<u_N\leq \ldots \leq u_1<2u_N, \hspace{.1in} u_1\not =u_N$. We have the following result.

\noindent
\textbf{Proposition:}
For any one-shot KPR game $\bar{G}(u)$, $SE(\bar{G}(u))=PN(\bar{G}(u))$.

\noindent {\bf Proof:} Consider any one-shot KPR game and a strategy combination $s^*\in SE(\bar{G}(u))$. Take any agent $i\in \mathcal{N}$ and any deviation $s_i=k\not =s^*_i=k'$. Since the number of agents is equal to the number of restaurants and since for a strategy combination $s^*\in SE(\bar{G}(u))$, $n_i(s^*)=1$ for all $i\in \mathcal{N}$, it follows that $n_i(s_i,s^*_{-i})=2>n_i(s^*)=1$. Therefore, $\Pi_i(s_i,s^*_{-i})=\frac{u_k}{2}<u_{k'}=\Pi_i(s^*)$ since for the one-shot KPR game $0<u_N\leq \ldots \leq u_1<2u_N\Leftrightarrow \frac{u_k}{2}<u_{k'}$ for all $k,k'\in \{1,\ldots,N\}$ and hence agent $i$ has no incentive to deviate. Since the selection of $i$ was arbitrary it follows that $s^*\in PN(\bar{G}(u))$ implying $SE(\bar{G}(u))\subseteq PN(\bar{G}(u))$.

To prove $PN(\bar{G}(u))\subseteq SE(\bar{G}(u))$ consider any strategy combination $s_{-i}$ for all but player $i$. Given $s_{-i}$, the best response of agent $i$ is to select a best ranked restaurant from the set $\mathcal{UN}_i(s_{-i})=\{k\in \mathcal{N}\mid s_j\not =k$ $\forall$ $j\not =i\}$ of unoccupied restaurants. The reasons being (1) given $\frac{u_1}{2}<u_N$, $\frac{u_k}{q}<u_{k'}$ for all $k,k'\in \{1,\ldots,N\}$ and all integers $q\geq 2$ implying that given the strategy of all other agents, agent $i$ strictly prefers to occupy a  best unoccupied restaurant as opposed to either occupying a different unoccupied restaurant (which is not a best restaurant) or crowding in an occupied restaurant and (2) given the number of agents is equal to the number of restaurants, $\mathcal{UN}_i(s_{-i})\not =\emptyset$ for any $s_{-i}\in S^{\mathcal{N}-1}$ implying that given any strategy of all other agents, agent $i$ can always find an unoccupied restaurant and hence agent $i$ can also find a best unoccupied restaurant since $0<u_N\leq \ldots \leq u_1<2u_N$. Therefore, it is in the interest of each agent to occupy an unoccupied restaurant and given the fact that no matter what others do an agent can always find an unoccupied restaurant it follows that any pure strategy Nash equilibrium for the one-shot KPR game necessarily leads to a socially efficient allocation, that is $PN(\bar{G}(u))\subseteq SE(\bar{G}(u))$.   

\end{document}